\newif\ifreport
\reportfalse

\documentclass[runningheads]{llncs}
\usepackage{graphicx}

\usepackage{amsthm}
\usepackage{amssymb}
\usepackage{amsmath}
\usepackage{paralist} \usepackage{caption}

\usepackage{hyperref}

\title{Offline Constrained\\ Backward Time Travel Planning}
\author{Quentin Bramas\inst{1} \and 
Jean-Romain Luttringer\inst{1} \and
Sébastien Tixeuil\inst{2,3}}

\authorrunning{Q. Bramas et al.}
\institute{ICUBE, Strasbourg University, CNRS, Strasbourg, France \and
Sorbonne University, CNRS, LIP6, Paris, France \and
Institut Universitaire de France, Paris, France}

\newcommand{\src}{\mathit{src}}
\newcommand{\dst}{\mathit{dst}}

\usepackage{etoolbox}
\newcommand{\contentFromMainDocument}{}
\newcommand{\movetoappendice}[2]{
\ifthenelse{\equal{#1}{\string no}}
{#2}
{\appto\contentFromMainDocument{
    #2
    }
}
}

\usepackage{proof-at-the-end}

\ifreport
    \pgfkeys{/prAtEnd/conf/.style={normal}
    }
    \pgfkeys{/prAtEnd/confall/.style={normal}
    }
    
    \RenewDocumentEnvironment{textAtEnd}{O{}+b}{#2}{}
\else
    \pgfkeys{/prAtEnd/conf/.style={end, restate, no link to proof}
    }
    \pgfkeys{/prAtEnd/confall/.style={all end
}}
\fi

\newcommand{\neig}{N}
 
\usepackage{multicol}
\usepackage{comment}

\newcommand{\myparagraph}[1]{\vspace{0.2cm}\noindent\textbf{#1.}}

\usepackage[vlined]{algorithm2e}

\SetCommentSty{mycommfont}

\usepackage{tikz}
\usepackage{numerica}

\usepackage{bitset}
\usepackage{ifthen}
\newcommand{\grid}[3]{
    \bitsetSetBin{mybitset}{#3}
    \foreach \griditer in {1,2,...,#2}
    {
        \ifthenelse{\bitsetGet{mybitset}{#2-\griditer}=1}
        {
            \pgfmathsetmacro{\griditerminusone}{int(\griditer-1)}
            \draw (\griditerminusone-#1) -- (\griditer-#1);
        }{
        }
    }
}

\newcommand{\initGrid}[2]{

\foreach \x in {0,1,...,#1} {
    \node at (\x, #2 + 0.5) {$x_\x$};
    \foreach \y in {0,1,...,#2} {
        \node[sty_node] (\x-\y) at (\x, #2-\y) {};
    }
}
\draw (0,#2+1) edge[-latex, thick] (#1, #2+1);
\node at (#1 / 2, #2+1.5) {\textbf{space}};

\node[] at (-1, #2) {$0$};
\foreach \y in {1,2,...,#2} {
    \node[] at (-1, #2-\y) {$\y$};
}
\draw (-2, #2) edge[-latex, thick] (-2, 0);
\node at (-2.7,#2 / 2) {\textbf{time}};
}

\newcommand{\ie}{\emph{i.e.}}
\newcommand{\N}{\mathbb{N}}
\newcommand{\Z}{\mathbb{Z}}
\newcommand{\R}{\mathbb{R}}

\newcommand{\F}{\mathcal{F}}

\newcommand\fcost{\mathfrak{f}}\newcommand\jcost{\mathit{cost}}
\newcommand\jdelay{\mathit{delay}}
\newcommand{\Const}{\mathcal{C}}
\newcommand{\Hmax}{\mathcal{H}}
\newcommand{\tmin}{t_{\min}}

\newcommand{\ST}[1]{\noindent\textcolor{olive}{{\fontfamily{phv}\selectfont ST-NOTE: #1}}}
\newcommand{\QB}[1]{\noindent\textcolor{red}{{\fontfamily{phv}\selectfont QB-NOTE: #1}}}
\newcommand{\JR}[1]{\noindent\textcolor{orange}{{\fontfamily{phv}\selectfont JR-NOTE: #1}}}

\iftrue
\renewcommand{\ST}[1]{}
\renewcommand{\QB}[1]{}
\renewcommand{\JR}[1]{}

\fi

\begin{document}

\maketitle

\begin{center}
    This version of the article has been accepted for publication in the proceedings of the International Symposium on Stabilizing, Safety, and Security of Distributed Systems (SSS 2023), after peer review and is subject to Springer Nature’s AM terms of use, but is not the Version of Record and does not reflect post-acceptance improvements, or any corrections. The Version of Record is available online at:\\
    \url{https://doi.org/10.1007/978-3-031-44274-2_35}
\end{center}

\begin{abstract}
We model transportation networks as dynamic graphs and introduce the ability for agents to use Backward Time-Travel (BTT) devices at any node to travel back in time, subject to certain constraints and fees, before resuming their journey.

We propose exact algorithms to compute travel plans with constraints on BTT cost or the maximum time that can be traveled back while minimizing travel delay (the difference between arrival and starting times). These algorithms run in polynomial time. We also study the impact of BTT device pricing policies on the computation of travel plans with respect to delay and cost and identify necessary properties for pricing policies to enable such computation. 
\end{abstract}

\let\oldsection\section

\section{Introduction}

Evolving graphs (and their many variants) are graphs that change over time and are used to model real-world systems that evolve. They have applications in many fields in Computer Science, where they arise in areas such as compilers, databases, fault-tolerance, artificial intelligence, and computer networks. 
To date, such graphs were studied under the hypothesis that time can be traveled in a single direction (to the future, by an action called waiting), leading to numerous algorithms that revisit static graph notions and results.

In this paper, we introduce the possibility of Backward time travel (BTT) (that is, the ability to go back in time) when designing algorithms for dynamic graphs. 
In more details, we consider the application of BTT devices to transportation networks modeled by evolving graphs. In particular we focus on the ability to travel from point $A$ to point $B$ with minimal delay (that is, minimizing the time difference between arrival and start instants), taking into account meaningful constraints, such as the cost induced by BTT devices, or their span (how far back in time you are allowed to go).

To this paper, BTT was mostly envisioned in simple settings (with respect to the cost associated to time travel or its span). For example, the AE model~\cite{russo2019} considers that a single cost unit permits to travel arbitrarily in both space and time, trivializing the space-time travel problem entirely. Slightly more constrained models such as TM~\cite{pal1960} and BTTF~\cite{zemeckis1985} consider devices that either: \emph{(i)} only permit time travel~\cite{pal1960} (but remain at the same position), or\emph{(ii)} permit either time travel or space travel, but not both at the same time~\cite{zemeckis1985}. However, the cost involved is either null~\cite{pal1960}, or a single cost unit per time travel~\cite{zemeckis1985}.

Instead, we propose to discuss BTT in a cost-aware, span-aware context, that implies efficiently using BTT devices within a transportation system (from a simultaneous delay and cost point of view), and the computation of the corresponding multi-modal paths.
More precisely, in this paper, we address the problem of space-time travel planning, taking into account both the travel delay of the itinerary and the cost policy of BTT device providers. The context we consider is that of transportation systems, where BTT devices are always available to the agents traveling. Using each BTT device has nevertheless a cost, decided by the BTT device provider, and may depend on the span of the backward time jump.  Although BTT devices are always active, the ability to go from one location to another (that is, from one BTT device to another) varies across time. 
We consider that this ability is conveniently modeled by a dynamic graph, whose nodes represent BTT devices, and whose edges represent the possibility to instantly go from one BTT device to another. 
Given a dynamic graph, we aim at computing travel plans, from one BTT device to another (the closest to the agent's actual destination), considering not only travel delay and induced cost, but also schedule availability and common limitations of BTT devices.

In the following, we study the feasibility of finding such travel plans, depending on the pricing policy. It turns out that when the schedule of connections is available (that is, the dynamic graph is known), very loose conditions on the pricing policy enable to devise optimal algorithms (with respect to the travel delay and induced cost) in polynomial time, given a cost constraint for the agents, or a span constraint for the BTT devices. 

\myparagraph{Related Work}
Space-Time routing has been studied, but assuming only forward time travel, \ie, waiting, is available. The idea of using dynamic graphs to model transportation network was used by many studies (see \emph{e.g.}~Casteigts et al.\cite{casteigts12peds} and references herein), leading to recently revisit popular problems previously studied in static graphs~\cite{casteigts15fcs,casteigts21css,luna21icdcs}. In a dynamic (or temporal) graph, a journey represents a temporal path consisting in a sequence of edges used at given non-decreasing time instants. The solvability of a problem can depend on whether or not a journey can contain consecutive edges occurring at the same time instant. Such journeys are called \emph{non-strict}, as opposed to \emph{strict} journey where the sequence of time instants must be strictly increasing. In our work, we extend the notion of non-strict journey to take into account the possibility to go back in time at each node, but one can observe that our algorithm also work with the same extension for strict journey by adding one time unit to the arrival of each edge in our algorithms.

The closest work in this research path is due to Casteigts et al~\cite{casteigts21algo}, who study the possibility of discovering a temporal path between two nodes in a dynamic network with a waiting time constraint: at each step, the traveling agent cannot wait more than $c$ time instants, where $c$ is a given constant. 
It turns out that finding the earliest arriving such temporal path can be done in polynomial time.
Perhaps surprisingly, Villacis-Llobet et al~\cite{villacis-llobet22sirocco} showed that if one allows to go several times through the same node, the obtained temporal path can arrive earlier, and finding it can be done in linear time. 
As previously mentioned, this line of work only considers \emph{forward} time travel: a temporal path cannot go back in time.

Constrained-shortest-paths computation problems have been extensively studied in the context of static graphs~\cite{QoS}. Although these problems tend to be NP-Hard~\cite{GareyJohnson1990} (even when considering a single metric), the ones considering two additive metrics (commonly, the delay and a cost) gained a lot of traction over the years due to their practical relevance, the most common use-case being computer networks~\cite{Garroppo_Giordano_Tavanti_2010,Guck_Van_Bemten_Reisslein_Kellerer_2018}. In this context, each edge is characterized by a weight vector, comprising both cost and delay. Path computation algorithms thus have to maintain and explore all non-comparable paths, whose number may grow exponentially with respect to the size of the network. To avoid a worst-case exponential complexity, most practical algorithms rely on either approximation schemes~\cite{handbook} or heuristics. However, these contributions do not study multi-criteria path computation problems within a time travel context. Conversely, we study and provide results regarding the most relevant time-traveling problems while considering the peculiarities of this context (in particular, the properties of the cost function). In  addition, we show that most of these problems can be solved optimally in polynomial-time.

\myparagraph{Contributions}
In this paper, we provide the following contributions:
\begin{itemize}\item An in-depth analysis of the impact of the BTT device providers pricing policies 
    on the computation of low-latency and low-cost paths. In particular, we show that few features are required to ensure that the efficient computation of such paths remains possible.
    \item Two exact polynomial algorithms able to compute travels with smallest delay to a given destination and minimizing the cost of traveling back in time. The first algorithm also supports the addition of a constraint on the backward cost of the solution. The other one supports a constraint on how far back in the past one can go at each given time instant.
\end{itemize}

\section{Model}

In this section, we define the models and notations used throughout this paper, before formalizing the aforementioned problems.

We represent the network as an evolving graph, as introduced by Ferreira~\cite{ferreira2002algotel}: a graph-centric view of the network that maps a dynamic graph as a sequence of static graphs. The \emph{footprint} of the dynamic graph (that includes all nodes and edges that appear at least once during the lifetime of the dynamic graph), is fixed. Furthermore, we assume that the set of nodes is fixed over time, while the set of edges evolves. 

More precisely, an evolving graph $G$ is a pair $(V, (E_t)_{t\in \N})$, where $V$ denotes the finite set of vertices, $\N$ is the infinite set of time instants, and for each $t\in \N$, $E_t$ denotes the set of edges that appears at time $t$. 
The \emph{snapshot} of $G$ at time $t$ is the static graph $G(t) = (V, E_t)$, which corresponds to the state, supposedly fixed, of the network in the time interval $t, t+1)$.
The \emph{footprint} $\F(G)$ of $G$ is the static graph corresponding to the union of all its snapshots, $\F(G) = \left(V, \bigcup_{t\in \N} E_t\right)$. We say $((u,v), t)$ is a temporal edge of graph $G$ if $(u,v)\in E_t$. We say that an evolving graph is \emph{connected} if its footprint is connected.

\myparagraph{Space-time Travel}
We assume that at each time instant, an agent can travel along any number of adjacent consecutive communication links. However, the graph may not be connected at each time instant, hence it may be that the only way to reach a particular destination node is to travel forward (\ie, wait) or backward in time, to reach a time instant where an adjacent communication link exists. In more detail, an agent travels from a node $s$ to a node $d$ using a \emph{space-time travel} (or simply travel when it is clear from the context). 
\begin{definition}
A \emph{space-time travel} of \emph{length} $k$ is a sequence $((u_0, t_0), (u_1, t_1),$ $\ldots, (u_k, t_k))$ such that
\begin{itemize}
    \item $\forall i\in\{0, \ldots k\}$, $u_i\in V$ is a node and $t_i \in \N$ is a time instant,
    \item $\forall i\in\{0, \ldots k-1\}$, if $u_i\neq u_{i+1}$, then $t_{i} = t_{i+1}$ and $(u_i, u_{i+1}) \in E_{t_i}$ \ie, there is a temporal edge between $u_i$ and $u_{i+1}$ at time $t_i$.
\end{itemize}
\end{definition}
By extension, the \emph{footprint} of a travel is the static graph containing all edges (and their adjacent nodes) appearing in the travel. Now, the \emph{itinerary} of a travel $((u_0, t_0), (u_1, t_1), \ldots, (u_k, t_k))$ is its projection $(u_0, u_1, \ldots, u_k)$ on nodes, while its \emph{schedule} is its projection $(t_0, t_1, \ldots, t_k)$ on time instants.

\begin{definition}
A travel $((u_0, t_0), (u_1, t_1), \ldots, (u_k, t_k))$ is \emph{simple} if for all $i \in\{2,\ldots, k\}$ and $j\in \{0, \ldots, i-2\}$, we have $u_i \neq u_j$.
\end{definition}

Intuitively, a travel is simple if its footprint is a line (\ie, a simple path) and contains at most one time travel per node (as a consequence, no node appears three times consecutively in a simple travel).

\begin{definition}
The \emph{delay} of a travel $T = ((u_0, t_0), (u_1, t_1), \ldots, (u_k, t_k))$, denoted $\jdelay(T)$ is defined as $t_k - t_0$.
\end{definition}

\myparagraph{The Backward cost of a travel}
\begin{definition}
The \emph{backward-cost} is the cost of going to the past. The backward-cost function $\fcost: \N^*\rightarrow \R^+$ returns, for each $\delta\in \N$, the backward-cost $\fcost(\delta)$ of traveling $\delta$ time instants to the past. As we assume that there is no cost associated to forward time travel (that is, waiting), we extend $\fcost$ to $\Z$ by setting $\fcost(-\delta) = 0$, for all $\delta\in\N$. In particular, the backward-cost of traveling $0$ time instants in the past is zero. When it is clear from context, the backward-cost function is simply called the cost function.
\end{definition}

\begin{definition}
The \emph{backward-cost} (or simply cost) of a travel $T = ((u_0, t_0)$, $(u_1, t_1)$, $\ldots, (u_k,$  $t_k))$, denoted $\jcost(T)$ is defined as follows:
\[
\jcost(T) = \sum_{i = 0}^{k-1} \fcost(t_i - t_{i+1})
\]
\end{definition}

\begin{definition}
Let $T_1 = ((u_0, t_0), (u_1, t_1), \ldots, (u_k, t_k))$ and $T_2=((u_0', t_0')$,$(u_1', t_1')$, $\ldots, (u_{k'}',$ $t_{k'}'))$ be two travels. If $(u_k,t_k) = (u_0', t_0')$, then the \emph{concatenated travel} $T_1\oplus T_2$ is defined as follows:
\[
 T_1\oplus T_2 = ((u_0, t_0), (u_1, t_1), \ldots, (u_k, t_k), (u_1', t_1'), \ldots, (u_{k'}', t_{k'}'))
\]
\end{definition}
\begin{remark}\label{rem:cost respect the addition}
One can easily prove that $\jcost(T_1 \oplus T_2) = \jcost(T_1) + \jcost(T_2)$. In the following, we sometimes decompose a travel highlighting an intermediate node: $T = T_1 \oplus ((u_i, t_i)) \oplus T_2$. Following the definition, this means that $T_1$ ends with $(u_i, t_i)$, and $T_2$ starts with $(u_i, t_i)$, so we also have $T = T_1 \oplus T_2$ and $\jcost(T) = \jcost(T_1) + \jcost(T_2)$. \end{remark}

Our notion of space-time travel differs from the classical notion of \emph{journey} found in literature related to dynamic graphs~\cite{ferreira2002algotel} as we do \emph{not} assume time instants monotonically increase along a travel. As a consequence, some evolving graphs may not allow a journey from $A$ to $B$ yet allows one or several travels from $A$ to $B$ (See Figure~\ref{fig:network2}). 

We say a travel is cost-optimal if there does not exist a travel with the same departure and arrival node and times as $T$ having a smaller cost. One can easily prove the following Property.
\begin{property}\label{lem:sub-travel is also cost-optimal}
Let $T$ be a cost-optimal travel from node $u$ to node $v$ arriving at time $t$, and $T'$ a sub-travel of $T$ \ie, a travel such that $T = T_1 \oplus T' \oplus T_2$. Then $T'$ is also cost-optimal. However, this is not true for delay-optimal travels.
\end{property}

\newcommand{\pbname}{ODOC\xspace}
\newcommand{\cpbname}{$\Const$-cost-constrained \pbname}
\myparagraph{Problem specification}
We now present the problems that we aim to solve in this paper. First, we want to arrive at the destination as early as possible, \ie, finding a time travel that minimizes the delay. Among such travels, we want to find one that minimizes the backward cost. 

In the remaining of this paper, we consider a given evolving graph $G=\left(V,(E_t)_{t\in\N}\right)$, a given a cost function $\fcost$, a source node $\src$ and a destination node $\dst$ in $V$. $\mathit{Travels}(G, \src, \dst)$ denotes the set of travels in $G$ starting from $\src$ at time $0$ and arriving at $\dst$.

\begin{definition}
The \emph{Optimal Delay Optimal Cost space-time travel planning} (\pbname) problem consists in finding, among all travels in $Travels(G, \src, \dst)$, the ones that minimize the travel delay and, among them, minimize the cost. A solution to the \pbname problem is called an \pbname travel.
\end{definition}

One can notice that this problem is not very hard as there is a single metric (the cost) to optimize, because a travel with delay zero always exists (if the graph is temporally connected). But in this paper we study the two variants defined thereafter (see the difference in bold).

\begin{definition}
The \textbf{$\Const$-cost-constrained} \pbname problem consists in finding a\-mong all travels in $\mathit{Travels}(G, \src, \dst)$ \textbf{with cost at most $\Const\geq 0$}, the ones that minimize the travel delay and, among them, one that minimizes the cost. \end{definition}
\begin{definition}
The \textbf{$\Hmax$-history-constrained} \pbname problem consists in finding among all travels in $\mathit{Travels}(G, \src, \dst)$ \textbf{satisfying,} $$
\forall u, u', t, t', \text{ if }T = T_1 \oplus ((u,t)) \oplus T_2 \oplus  ((u',t'))\oplus T_3, \text{ then } t'\geq t - \Hmax,
$$
the ones that minimize the travel delay and, among them, one that minimizes the cost. \end{definition}

\tikzstyle{sty_node}=[circle, draw, thin,fill=black!20, inner sep=0.07cm]

\noindent\begin{minipage}[t]{0.47\textwidth}
\centering
\hspace*{-1cm}\begin{tikzpicture}[scale=0.5]
\initGrid{7}{7}

\grid{0}{7}{1000101}
\grid{1}{7}{0101011}
\grid{2}{7}{1100100}
\grid{3}{7}{0110111}
\grid{4}{7}{1010011}
\grid{5}{7}{1111011}
\grid{6}{7}{1010110}
\grid{7}{7}{0101011}

\draw [black] plot [smooth] coordinates {(3,7) (4,6.7) (5,7)};
\draw [black] plot [smooth] coordinates {(3,5) (4,4.7) (5,5)};
\draw [black] plot [smooth] coordinates {(1,5) (2,4.7) (3,5)};
\draw [black] plot [smooth] coordinates {(0,4) (1,3.7) (2,4)};
\draw [black] plot [smooth] coordinates {(4,4) (5,3.7) (6,4)};
\draw [black] plot [smooth] coordinates {(0,3) (1,2.7) (2,3)};
\draw [black] plot [smooth] coordinates {(5,3) (6,2.7) (7,3)};
\draw [black] plot [smooth] coordinates {(2,3) (3.5,2.6) (5,3)};
\draw [black] plot [smooth] coordinates {(2,2) (3,1.7) (4,2)};
\draw [black] plot [smooth] coordinates {(1,1) (2,0.7) (3,1)};
\draw [black] plot [smooth] coordinates {(3,1) (4,0.7) (5,1)};
\draw [black] plot [smooth] coordinates {(0,0) (1,-0.3) (2,0)};
\draw [black] plot [smooth] coordinates {(2,0) (3,-0.3) (4,0)};
\draw [black] plot [smooth] coordinates {(4,0) (5,-0.3) (6,0)};

\draw [red,line width=0.25mm] plot [smooth] coordinates {(0,7) (1,7) (1,6) (2,6) (2,4) (3,4) (3,2) (4,2) (4,1) (6,1) (6,0) (7,0)};

\draw [blue,line width=0.25mm] plot [smooth] coordinates {(0,6.95) (1,6.95) (1,5.95) (1.95,5.95) (1.95,3.95) (3,3.95) (3,5) (4,4.55) (5,5) (5,4) (7,4) (7.3,5.6) (7,7)};

\draw [green!80!black,line width=0.25mm] plot [smooth] coordinates {(0,7.05) (1,7.05) (1,6.05) (1.95,5.95) (2.05,4.05) (3,4.05) (3,5.05) (4,4.51) (5,5.05) (5,6) (7,6) (7,7)};

\end{tikzpicture}
\captionof{figure}{Possible representation of an evolving graph. Possible travels from $x_0$ to $x_7$ are shown in red, green and blue. Note that the blue and green travels require to send an agent to the past (to a previous time instant).}
\label{fig:network}
\end{minipage}\hfill
\begin{minipage}[h]{0.47\textwidth}
\centering
\begin{tikzpicture}[scale=0.45]
\draw node at (105:22.5) {};
    
    \foreach[count=\i from 0] \fillc in {white,white,white,white,white,white,white,white} {
\node[minimum size=4mm,inner sep=0,circle,draw,fill=\fillc] (point\i) at ({-\i/80*360+105}:20) {$x_\i$};
}
    
    \draw (point0)-- (point1);
    \draw (point0) edge[bend right] (point2);
    \draw (point1)-- (point2);
    \draw (point1) edge[bend right]  (point3);
    \draw (point2)-- (point3);
    \draw (point2) edge[bend right]  (point4);
    \draw (point2) edge[bend right]  (point5);
    \draw (point3)-- (point4);
    \draw (point3) edge[bend right]  (point5);
    \draw (point4)-- (point5);
    \draw (point4) edge[bend right]  (point6);
    \draw (point5)-- (point6);
    \draw (point5) edge[bend right]  (point7);
    \draw (point6)-- (point7);
\end{tikzpicture}
\captionof{figure}{Footprint of the evolving graph represented in Figure~\ref{fig:network}.}
\label{fig:network-footprint}
\end{minipage}

\noindent\begin{minipage}{0.47\textwidth}
\centering
\hspace*{-1cm}\begin{tikzpicture}[scale=0.5]
\initGrid{7}{2}

\grid{0}{7}{1010100}
\grid{1}{7}{0101010}
\grid{2}{7}{1001101}

\draw [black] plot [smooth] coordinates {(1,0) (3,-0.3) (5,0)};

\draw [blue,line width=0.25mm] plot [smooth] coordinates {(0.05,2.05) (1.05,2.05) (1.05,1.05) (1.95,0.8) (2.05,2.05) (3.05, 2.05) (3.05, 0.95) (4.05,1.0) (4.05,2.05) (5.05, 2.05) (5.05, 0.95) (6.05,1.0) (6,0.05) (7,0) (7,1)
};

\draw [green!80!black!80!black,line width=0.25mm] plot [smooth] coordinates {(0.1,2) (1.1,2) (1.1,0) (3.1,-0.3) (5.1,-0.1) (5.1,-0.1) (5.1,1) (6.1,1) (6.1,0) (7.1,0) (7.1,1)};

\end{tikzpicture}
\captionof{figure}{Example of an evolving graph for which there exists no journey, yet there exists several travels from $x_0$ to $x_7$. The two travels, in blue and green, are 1-history-constrained.}
\label{fig:network2}
\end{minipage}\hfill
\begin{minipage}{0.47\textwidth}
\centering
\begin{tikzpicture}[scale=0.5]
\initGrid{7}{3}

\grid{0}{7}{1010101}
\grid{1}{7}{0101011}
\grid{2}{7}{0111100}
\grid{3}{7}{1011011}

\draw [red, line width=0.25mm] plot [smooth] coordinates {(0,3) (1,3) (1,2) (2,2) (2,1) (5,1) (5,0) (7,0) (7,1)};

\draw [blue, line width=0.25mm] plot [smooth] coordinates {(0.05,3) (1.05,3) (1.05,2) (2.05,2) (2.05,1) (5.05,1) (5.05,2) (7.05,2)};

\draw [green!80!black, line width=0.25mm] plot [smooth] coordinates {(0,3.05) (1,3.05) (1,2.05) (2,2.05) (2,3.05) (3,3.05) (3,2.05) (4,2.05) (4,1.05) (5,1.05) (5,0.05) (7,0.05)};

\end{tikzpicture}

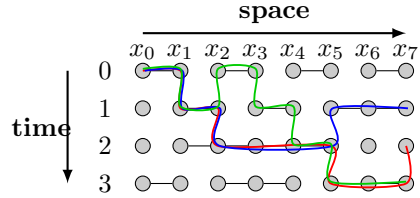
\captionof{figure}{Example of an evolving graph for which there exist at least three travels from $x_0$ to $x_7$ with a cost constraint of 1 (assuming $\fcost: d \mapsto d$). The blue travel has optimal delay.}
\label{fig:network3}
\end{minipage}

\myparagraph{Visual representation of space-time travels.}
To help visualize the problem, consider a set of $n+1$ nodes denoted $x_0, x_1, x_2, \ldots, x_n$. Then, the associated evolving graph can be seen as a vertical sequence of graphs mentioning for each time instant which edges are present.
A possible visual representation of an evolving graph can be seen in Figure~\ref{fig:network}. One can see the evolution of the topology (consisting of the nodes $x_0$ to $x_7$) over time through eight snapshots performed from time instants $0$ to $7$. Several possible travels are shown in red, green and blue. 
The red travel only makes use of forward time travel (that is, waiting) and is the earliest arriving travel in this class (arriving at time $7$).
The green and blue travels both make use of backward time travel and arrive at time 0, so they have minimal travel delay. Similarly, the red travel concatenated with $((x_7,7), (x_7, 0))$ (\ie, a backward travel to reach $x_7$ at time 0) also has minimal travel delay.
However, if we assume that the cost function is the identity ($\fcost: d \mapsto d$) then the green travel has a backward cost of 3, the blue travel has a backward cost of 4, and the concatenated red travel has a backward cost of 7. 
Adding constraints yields more challenging issues: assuming $\fcost: d \mapsto d$ and a maximal cost $\mathcal{C}$ of $1$, at least three travels can be envision for the evolving graph depicted in Figure~\ref{fig:network3}, but finding the $1$-cost-constrained travel that minimizes the delay (that is, the blue travel) is not as straightforward in this case, even if the footprint of the evolving graph is a line.

Similarly, in Figure~\ref{fig:network2} we show two $\Hmax$-history-constrained travels, with $\Hmax = 1$ (assuming $\fcost: d \mapsto d$). Here, clearly, the green travel is optimal with a cost of $2$ (the blue travel has cost $3$). The choice made by the green travel to wait at node $x_1$ two time instants is good, even if it prevents future backward travel to time 0 since $\Hmax=1$; because it is impossible to terminates at time 0 anyway. So it seems like the choice made at node $x_1$ is difficult to make before knowing what is the best possible travel. If we add more nodes to the graph and repeat this kind of choice, we can create a graph with an exponential number of $1$-history-constraint travel and finding one that minimizes the cost is challenging. Surprisingly, we show that it remains polynomial in the number of nodes and edges.

\section{Backward-cost Function Classes}

The cost function $\fcost$ represents the cost of going back to the past. 
Intuitively, it seems reasonable that the function is non-decreasing (travelers are charged more it they go further back in time), however we demonstrate that such an assumption is not necessary to enable travelers to derive optimal cost space-time travel plans. 
As a matter of fact, the two necessary conditions we identify to optimally solve the \pbname space-time travel planning problem are $\fcost$ to be non-negative and that it attains its minimum (not just converge to it). These conditions are shown to be sufficient by construction, thanks to the algorithm presented in the next section (and Theorem~\ref{thm:algorithm assuming user friendly functions can be transformed}). Due to space constrains, proofs are omitted.

\begin{definition}
A cost function $\fcost$ is \emph{user optimizable} if it is non-negative, and it attains its minimum when restricted to any interval $[C,\infty)$, with $C>0$.
Let $\mathcal{UO}$ be the set of user optimizable cost functions.
\end{definition}

\begin{theoremEnd}[normal]{theorem}
If the cost function $\fcost$ is not in $\mathcal{UO}$, then there exist connected evolving graphs where no solution exists for the \pbname space-time travel planning problem.
\end{theoremEnd}
\begin{proofEnd}
First, it is clear that if $\fcost(d) < 0$ for some $d\in \N^*$, then we can construct travels with arbitrarily small cost by repeatedly appending $((y,t),(y,t+d),(y,t))$ to any travel arriving at node $y$ at time $t$ (\ie, by waiting for $d$ rounds and going back in time $d$ rounds), rendering the problem unsolvable. 

Now, let $C\in\N^*$ and $\fcost$ be a non-negative function that does not attain its minimum when restricted to $[C,\infty)$. This implies that there exists an increasing sequence $(w_i)_{i\in \N}$ of integers $w_i\geq C$, such that the sequence $(\fcost(w_i))_{i\in \N}$ is decreasing and converges towards the lower bound $m_C = \text{inf}_{t\geq C}(\fcost(t))$ of $\fcost_{|_{[C,\infty)}}$. Consider a graph with two nodes $x_0$ and $x_1$ that are connected by a temporal edge after time $C$ and disconnected before. 
 Since a travel from $x_0$ to $x_1$ arriving at time $0$ must contain a backward travel to the past of amplitude at least $C$, its cost is at least equal to $m_C$. Since $m_C$ is not attained, there is no travel with cost exactly $m_C$. 
 Now, assume for the sake of contradiction that a cost-optimal travel $T$ to $x_1$ arriving at time $0$ has cost $m_C+\varepsilon$ with $\varepsilon>0$. Then, we can construct a travel with a smaller cost. Let $i_{\varepsilon}$ such that $\fcost(w_{i_{\varepsilon}}) < m_C + \varepsilon$ (this index exists because the sequence $(\fcost(w_i))_{i\in \N}$ converges to $m_C$).
 
 Let $T' = ((x_0, 0), (x_0, C), (x_1, C), (x_1, w_{i_{\varepsilon}}), (x_1, 0))$.
 Then we have \[
 \jcost(T') = \fcost(w_{i_{\varepsilon}})<m_C + \varepsilon = \jcost(T),
 \] which contradicts the optimality of $T$.
\end{proofEnd}

We now present the set of \emph{user friendly} cost functions that we use in the sequel to ease proving optimization algorithms, as they allow \emph{simple} solutions to the \pbname problem (Lemma~\ref{lem:simple solutions exists when function is user friendly}).
We prove in Theorem~\ref{thm:algorithm assuming user friendly functions can be transformed} that we do not lose generality since an algorithm solving the \pbname problem with user friendly cost functions can be transformed easily to work with any user optimizable ones.

\begin{definition}
A cost function $\fcost$ is \emph{user friendly} if it is user optimizable, non-decreasing, and sub-additive\footnote{sub-additive means that for all $a,b\in \N$, $\fcost(a+b) \leq \fcost(a) + \fcost(b)$}. 
Let $\mathcal{UF}$ be the set of user friendly cost functions.
\end{definition}

\begin{theoremEnd}[normal]{lemma}\label{lem:simple solutions exists when function is user friendly}
If the cost function $\fcost$ is in $\mathcal{UF}$ and there exists a solution to the \pbname space-time travel planning problem in an evolving graph $G$, 
then there also exists a \emph{simple} travel solution.
\end{theoremEnd}

\begin{proofEnd}
Let $T$ be a solution to the \pbname space-time travel planning problem.
If there exists a node $x_i$ and two time instants $t_1$ and $t_2$, such that $T = T_1 \oplus ((x_i, t_1)) \oplus T_2 \oplus ((x_i, t_2)) \oplus T_3$, then we construct $T'$ as follows
\[
T' =  T_1 \oplus ((x_i, t_1), (x_i, t_2)) \oplus T_3
\]
and we show that $\jcost(T') \leq \jcost(T)$. Indeed, it is enough to show (thanks to Remark~\ref{rem:cost respect the addition}) that \[
\jcost(((x_i, t_1), (x_i, t_2))) \leq \jcost(T_2).
\]

By definition $\jcost(((x_i, t_1), (x_i, t_2))) = \fcost(t_1 - t_2)$. If $t_1 < t_2$, then the cost is null by convention and the Lemma is proved. Otherwise $t_1 > t_2$. On the right hand side, we have:
\[
\jcost(T_2) = \sum_{i = 1}^k \fcost(d_i)
\]
where $d_1, d_2, \ldots, d_k$ is the sequence of differences between the times appearing in $T_2$. Since $T_2$ starts at time $t_1$ and ends at time $t_2$, then $\sum_{i = 1}^k d_i = t_1 - t_2$. Since the function is sub-additive and increasing, we obtain: 
\[
\fcost(t_1 - t_2) < \sum_{i = 1}^k \fcost(d_i) 
\]

By repeating the same procedure, we construct a time-travel with the same destination and same backward-cost as $T$ but that does not contain two occurrences of the same node, except if they are consecutive.
\end{proofEnd}

\newcommand{\ftild}{\widetilde{\fcost}}

\begin{theoremEnd}[normal]{theorem}\label{thm:algorithm assuming user friendly functions can be transformed}
If an algorithm $A$ solves the optimal cost space-time travel planning problem for any cost function in $\mathcal{UF}$, then there exists an algorithm $A'$ solving the same problem with any $\fcost$ in $\mathcal{UO}$.
\end{theoremEnd}

\begin{proofEnd}
We consider an algorithm $A$ as stated. Let $\fcost$ be an arbitrary cost function in $\mathcal{UO}$, that is, $\fcost$ is non-negative, and always attains its minimum.

From $\fcost$, we now construct a cost function $\fcost_{inc}$ as follows:

\[
\fcost_{inc}(t) = 
\min_{j\geq t}\left(\fcost(j)\right)
\]
By construction, $\fcost_{inc}$ is non-decreasing. 
Moreover, since $\fcost$ is in $\mathcal{UO}$, it always attains its minimum, and we have:
\begin{equation}\label{eq:dm exists from d}
    \forall d,\;\exists d_m\text{ such that }\fcost_{inc}(d)=\fcost(d_m).
\end{equation}

Then, we construct $\widetilde{\fcost}$ as follows:
\[
\ftild(t) =  \min_{a \in \alpha(t)}\left(\sum_{a_i\in a}\fcost_{inc}(a_i)\right)
\]
where $\alpha(t)$ is the set of all the non-negative sequences that sum to $t$.
Now, $\ftild$ is sub-additive by construction, hence $\ftild\in\mathcal{UF}$. Since $\alpha(t)$ is finite, the minimum is attained.

Also, $\forall t\geq 1$, $\ftild(t) \leq \fcost(t)$, so that for any travel, its backward cost with respect to $\fcost$ is at least equal to its backward cost with respect to $\ftild$.

Let $G$ be a dynamic graph. Our goal is to construct an algorithm $A'$ finding a cost-optimal (with respect to $\fcost$) space-time travel in $G$. The algorithm $A'$ works as follows. Let $\widetilde{T}$ be an optimal solution found by algorithm $A$ on $G$ assuming function $\ftild$ is used. $A'$ now constructs, from $\widetilde{T}$, a time-travel $T$ that is a cost-optimal (with respect to $\fcost$) on $G$.

The travel $T$ is constructed from $\widetilde{T}$ by replacing any sub-space-time travel $((x_i, t_i),(x_i, t_i - t))$, with $t\geq 0$, by the following sub space-time travel:  $((x_i, t_i-a_1),(x_i, t_i-a_1-a_2),\ldots,(x_i, t_i-\sum_{j=1}^k a_j))$
satisfying:
\begin{align*}
a\in\alpha(t) \quad\wedge\quad \ftild(t) = \sum_{j=1}^{\text{length of $a$}} \fcost_{inc}(a_j)
\end{align*}

Then, each $((u, t),(u, t-d))$, with $d\geq 0$, is replaced by $((u, t), (u, t-d+d_m),(u, t-d))$ such that:
\begin{align*}
d_m \geq d \quad\wedge\quad 
\fcost_{inc}(d_m) = \fcost(d)
\end{align*}
We know that $d_m$ exists thanks to Equation~\ref{eq:dm exists from d}.
The space-time travel $T$ uses the same temporal edges as $\widetilde{T}$, so it is well defined. Moreover, by construction $\fcost(T) = \ftild(\widetilde{T})$, and $T$ is optimal with respect to $\fcost$ because the backward-cost of a travel with respect to $\fcost$ is at least equal to its backward-cost with respect to $\ftild$, as observed earlier. Hence, if a better solution exists for $\fcost$, it is also a solution with the same, or smaller, cost with $\ftild$, contradicting the optimality of $\widetilde{T}$.
The above procedure defines an algorithm, based on $A$, that solves the \pbname problem with function $\fcost$.
\end{proofEnd}

\section{Offline \cpbname Algorithm}
\label{sec:offline_unlimited}

In this section, we present Algorithm~\ref{alg:offline} that solves the \cpbname problem in time polynomial in the number of edges. More precisely, since the number of edges can be infinite, we only consider edges occurring before a certain travel (see the end of the section for a more precise description of the complexity). Algorithm~\ref{alg:offline} is different from existing shortest path algorithms because we need to efficiently take into account the cost and the delay of travels. It is well-known that constrained shortest path algorithms are exponential when considering two additive metrics~\cite{WangCrowcroft1996} but surprisingly, our algorithm is polynomial by using the specificity of the time travel. Our algorithm works as follows. At each iteration, we extract the minimum cost to reach a particular node at a particular time and we extend travels from there by updating the best-known cost of the next node. We reach the next nodes either by using the next temporal edge that exists in the future (we prove that considering only the next future edge is enough) or using each of the past temporal edge.

\RestyleAlgo{ruled}
\LinesNumbered

\SetKwBlock{Forever}{forever}{end}

\newcommand{\pq}{\texttt{pq}}
\newcommand{\nodeCost}{\texttt{nodeCost}}
\newcommand{\minCost}{\texttt{minCost}}
\newcommand{\pred}{\texttt{pred}}
\newcommand{\push}{\ensuremath{\mathit{push}}}
\newcommand{\m}[1]{\ensuremath{\mathit{#1}}}
\newcommand{\done}{\texttt{done}}
\newcommand{\futt}{t_{\mathit{future}}}
\newcommand{\pastt}{t_{\mathit{past}}}
\newcommand{\pastc}{c_{\mathit{past}}}
\begin{algorithm}[t]
\small
\caption{Offline $\Const$-cost-constrained \pbname Algorithm (input: $G, \fcost, \Const, \src, \dst$)}\label{alg:offline}
\tcc{\nodeCost[u,t] stores the current best cost of travels from node $\src$ to node $u$ arriving at time $t$.\quad
$\minCost[u]$ stores a pair $(c,t)$ where $c$ is the current known minimum cost of a travel towards $u$, and $t$ the smallest time where such travel arrives.
$\pred[u, t]$ stores the suffix of an optimal travel to $u$ arriving at $t$. }
$\forall u\in V, \forall t, \quad \nodeCost[u,t] = \infty\quad \minCost[u] = (\infty, \infty)$\;
$\nodeCost[\src, 0] \leftarrow 0$; $\qquad\done \leftarrow \emptyset$\;\label{algo1:line init 0,0}

\While{\textnormal{$\exists (u, t)\notin \done$ such that $\nodeCost[u, t] < \infty$}}
{
    $(u, t) \leftarrow \text{argmin}_{(u, t)\notin \done}(\nodeCost[u, t])$ \label{line:offline:extract min}\;
    $\done \leftarrow \done\cup\{(u,t)\}$\;
    $c\leftarrow\nodeCost[u, t]$\;
    \For{each neighbor $v$ of $u$}{
        let $\futt$ the smallest time after (or equal to) $t$ where edge $((u, v),\futt)$ exists\label{algo1:line extract future 1}\;
        let $(c_{min}, t_{min}) = \minCost[v]$\;
        \If{\textnormal{$\nodeCost[v,\futt] > c$ and ($c < c_{min}$ or $\futt < t_{min}$)}\label{line:condition to append v,t_futt}}
        {
            
            $\nodeCost[v,\futt] \leftarrow c$\;\label{algo1:line extract future 2}
            $\pred[v,\futt] \leftarrow ((u,t), (u,\futt), (v,\futt))$\;
            
            \lIf{\textnormal{$(c, \futt) <_{lexico} \minCost[v]$}}{
                $\minCost[v]\leftarrow (c,\futt)$\label{algo1:line update minCost}
            }
        }
        
        \For{\textnormal{each $\pastt$ such that $(u, v)\in E_{\pastt}$\label{algo1:line extract past 1}}}{
            let $\pastc = c + \fcost(t-\pastt)$\;
            \If{\textnormal{$\pastc \leq \Const$ and $\nodeCost[v, \pastt] > \pastc$}\label{line:condition to append v,t_past}}
            {
                $\nodeCost[v, \pastt] \leftarrow \pastc$\;\label{algo1:line extract past 2}
               $\pred[v,\pastt] \leftarrow ((u,t), (u,\pastt), (v,\pastt))$\;
            } 
        }
    }
}
let $\tmin$ be the minimum time instant such that $\exists t$, $\nodeCost[\dst,t] + \fcost(t - \tmin)\leq \Const$\;
\leIf{$\tmin$ exists}
{
    \Return \texttt{ExtractTimeTravel}($\dst,\tmin, \nodeCost, \pred$)\;
}{
\Return $\bot$
}
\end{algorithm}

\begin{textAtEnd}[category=extract-prop-algo]

\begin{algorithm}[H]
\small
\SetAlgoRefName{ExtractTimeTravel}
\caption{Extract a $\Const$-cost-constrained \pbname travel to the given destination}
\SetKwInOut{Input}{input}
\Input{$u\in V,\quad t\in\N,\quad \nodeCost,\quad \pred$}
\tcc{if the cost is null, $u$ is the source}
\If{$\nodeCost[u,t] = 0$}{
\Return $((u,0), (u, t))$\\
}
\If{$\nodeCost[u,t] = \infty$}{
    Let $t' = argmin_{t'\in\N}\left( \nodeCost[u,t']+\fcost(t'-t)\right)$\label{line to comute t' when nodecost in infinite}\;
    \Return \texttt{ExtractTimeTravel}($u,t',\nodeCost, \pred$)$\oplus((u,t'),(u,t))$\\
}
$((v,t'), (v,t), (u,t)) = \pred[u,t]$\;
\Return \texttt{ExtractTimeTravel}($v,t',\nodeCost, \pred$)$\oplus ((v,t'), (v,t), (u,t))$\\
\end{algorithm}
\end{textAtEnd}

We first prove that our algorithm terminates, even if the graph is infinite and if there is no solution. 

\begin{theoremEnd}[normal]{lemma}\label{lem:algo 1 terminates}
Algorithm~\ref{alg:offline} always terminates.
\end{theoremEnd}
\begin{proofEnd}

Assume for the sake of contradiction that it does not terminates. First, we observe that, for any $u\in V$, $\minCost[u]$ is non-increasing (using the lexicographical order), so it must reach a minimum value $(c_{u,\min}, t_{u,\min})$, which represent, for a node $u$, the minimum cost a travel towards $u$ can have and the minimum time such a travel can arrive.
Moreover, the cost associated with a pair $(u,t)$ extracted in Line~\ref{line:offline:extract min} is non-decreasing (because we always extract a pair with minimum cost), so either this cost reach a maximum or tends to infinity.
In the former case, let $c_{\max}$ be that maximum \ie, after some time, every time a pair $(u,t)$ is extracted, $\nodeCost[u,t] = c_{\max}$. Since a pair is never extracted twice, pairs are extracted with arbitrarily large value $t$. Some, at some point in the execution, for every pair $(u,t)$ extracted, we have $t > t_{u,\min}$. Moreover, $c_{\max} \geq c_{u,\min}$. So, every time a pair is extracted, condition Line~\ref{line:condition to append v,t_futt} is false. Hence, $c_{\max}$ is not added into $\nodeCost$ anymore, which contradicts the fact that $c_{\max}$ is associated with each extracted pair after some time.
So the latter case occurs \ie, the cost associated with extracted pairs tends to infinity. After some time, this cost is greater than any $c_{u, \min}$. Again, since a pair is never extracted twice, pairs are extracted with arbitrarily large value $t$. Some, at some point in the execution, for every pair $(u,t)$ extracted, we have $t > t_{u,\min}$, and the condition Line~\ref{line:condition to append v,t_futt} is always false.
Hence, from there, every time a value is added into $\nodeCost$, it is according to Line~\ref{algo1:line extract past 2}, so the associated time smaller than the time extracted, which contradicts the fact that arbitrarily large value $t$ are added to $\nodeCost$.
\end{proofEnd}

We now prove the correctness of our algorithm, starting with the main property we then use to construct a solution. Let $\delta_{\Const}$ be the function that returns, for each pair $(u,t)$ where $u$ is a node and $t$ a time, the best backward-cost smaller or equal to $\Const$, from $\src$ to $u$, for travels arriving at time~$t$.
\begin{theoremEnd}[normal]{lemma}\label{lem:nodeCost is the optimal cost}
When a pair $(u, t)$ is extracted from $\nodeCost$ at line~\ref{line:offline:extract min}, then $$\delta_{\Const}(u,t) = \nodeCost[u,t]$$ 
\end{theoremEnd}
\begin{proofEnd}
Assume for the sake of contradiction that this is not true, and let $(u, t)$ be the first tuple extracted such that the property is false. Let $c_{u,t} = \nodeCost[u, t]$. Let $T$ be a $\Const$-cost-constrained backward-cost-optimal travel to $u$ arriving at time $t$ (hence $\jcost(T) < c_{u,t}$ by assumption).

Let $T'$ be the longest prefix of $T$, to $(x,t')$ (\ie, such that $T=T'\oplus(x,t')\oplus T''$, for some $T''$), such that $(x,t')$ was extracted from $\nodeCost$ and satisfies $\delta_{\Const}(x,t') = \nodeCost[x,t']$. 
Now, $T'$ is well defined because the first element in $T$ is $(\src,0)$ and, by Line~\ref{algo1:line init 0,0}, $(\src,0)$ is the first extracted pair, and satisfies $\nodeCost[\src,0]=0=\delta_{\Const}(\src,0)$. Hence, prefix $((\src, 0))$ satisfies the property, so the longest of such prefixes exists. Observe that $T'$, resp. $T''$, ends, resp. starts, with $(x,t')$, by the definition of travel concatenation.

When $(x,t')$ is extracted from $\nodeCost$, it is extended to the next future edge (Lines~\ref{algo1:line extract future 1} to \ref{algo1:line extract future 2}), and all past edges (Lines~\ref{algo1:line extract past 1} to \ref{algo1:line extract past 2}). $T''$ starts either $(a)$ with $((x,t'), (x, t_a), (y, t_a))$, with $t_a < t'$, $(b)$ with $((x,t'), (x, t_a), (y, t_a))$ with $t_a > t'$, or $(c)$ with $((x,t'),(y, t'))$, where $y\in \neig(x)$.

\textbf{In case $(a)$,} this means that the temporal edge $((x, y), t_a)$ exists, hence, by Line~\ref{algo1:line extract past 2}, we know that $\nodeCost[y, t_a]\leq\nodeCost[x, t'] + \fcost(t' - t_a)$.
However, since $T'$ is a sub-travel, $\jcost(T') = \delta_{\Const}(x,t') = \nodeCost[x, t']$, hence 
\[
\nodeCost[y, t_a]\leq \jcost(T'\oplus ((x,t'),(x, t_a),(y, t_a))) = \delta_{\Const}(y,t_a),
\]
and $(y,t_a)$ must have been extracted before $(u,t)$, otherwise 
\[
\delta_{\Const}(u,t)<\nodeCost[y, t]\leq \nodeCost[y, t_a] = \delta_{\Const}(y,t_a) 
\]
which is a contradiction (a sub-travel of a cost-optimal travel cannot have a greater cost, see Property~\ref{lem:sub-travel is also cost-optimal}).
So, $T'\oplus ((x,t'),(x, t_a),(y, t_a))$ is a longer prefix of $T$ with the same property as $T'$, 
which contradicts the definition of $T'$.

\textbf{In case $(b)$,} this means that the temporal edge $((x, y),t_a)$ exists, hence, by Line~\ref{algo1:line extract future 2}, we know that $\nodeCost[y,t_a]\leq \nodeCost[x,t']$. Again, we have $\jcost(T') = \delta_{\Const}(x,t') = \nodeCost[x, t']$, hence 
\[
\nodeCost[y,t_a] \leq \jcost(T'\oplus ((x,t') , (x,t_a), (y,t_a))) = \delta_{\Const}(y,t_a),
\] 
which contradicts the definition of $T'$.

\textbf{In case $(c)$,} this means that the edge $((x,y),t')$ exists, which implies, using a similar argument, a contradiction.
\end{proofEnd}

The previous lemma says that $\nodeCost$ contains correct information about the cost to reach a node, but actually, it does not contain all the information. Indeed, a node $u$ can be reachable by a travel at a given time $t$ and still $\nodeCost[u,t] = \infty$. This fact helps our algorithm to be efficient, as it does not compute all the optimal costs for each possible time (in this case, the complexity would depend on the duration of the graph, which could be much higher than the number of edges). Fortunately, we now prove that we can still find all existing travel using $\nodeCost$.

\begin{theoremEnd}[conf]{lemma}\label{lem:nodeCost can be used to find paths}
For all $u \in V$, $t\in\N$, there exists a  $\Const$-cost-constrained travel $T$ from $\src$ to $u$ arriving at time $t$, if and only if there exists $t'\in\N$ such that $\nodeCost[u, t'] + \fcost(t'-t) \leq \Const$. \end{theoremEnd}
\begin{proofEnd}
We just need to prove the implication since the converse follows from the previous Lemma. Indeed, if $\nodeCost[u, t'] + \fcost(t'-t) \leq \Const$, then $\nodeCost[u, t']$ is finite and is the optimal cost of $\Const$-cost-constrained travels to $u$ arriving at time $t'$, so a $\Const$-cost-constrained travel to $u$ arriving at time $t$ exists.

The proof of the implication is done by induction on the length of the travel $T$ towards $(u,t)$. The result is clearly true when $T=((\src, 0))$ because $\delta_\{\Const\}(\src, 0) = 0$,Now let $T$ be a  $\Const$-cost-constrained travel to $u\in V\setminus \{\src\}$ arriving at $t\in\N$ and assume the result true for any travel of length smaller than the length of $T$. Assume for the sake of contradiction that
\begin{equation}\label{eq:the contradiction with cost[i,t'] + fcost(t'-t)}
    \text{$\forall t'\in \N$, $\nodeCost[u, t'] + \fcost(t'-t) > \Const$.}
\end{equation} 
We can assume w.l.o.g that $T$ is simple and cost optimal. If $T$ actually arrives at $u$ at time $t_b < t$, then we can use the inductive hypothesis. There is a shorter travel to $(u,t_b)$ so there exists $t'$ such that $\nodeCost[u, t']+ \fcost(t'-t_b)\leq \Const$, but since $t_b < t$ we have 
\[
 \nodeCost[u, t']+ \fcost(t'-t)\leq  \nodeCost[u, t']+ \fcost(t'-t_b) \leq \Const
\]
which contradicts Equation~(\ref{eq:the contradiction with cost[i,t'] + fcost(t'-t)}).

Hence $T$ arrives at $u$ through a temporal edge $((x,u), t_x)$ with $t_{x}\geq t$:
\[
T = T_1\oplus ((x, t_{x}), (u, t_{x}),(u, t))
\]
By inductive hypothesis, there exists $t'\in \N$ such that \begin{equation}\label{eq:inductive hyp on T'}
    \nodeCost[x, t'] + \fcost(t'-t_{x}) \leq \Const
\end{equation}

\textbf{If $t' < t_{x}$, then,} when the pair $(x, t')$ was extracted from $\nodeCost$ (in Line~\ref{line:offline:extract min}), since an edge exists between $x$ and $u$ at time $t_{x}$, then the variable $\futt$ is at most $t_{x}$ and we have
\[
\nodeCost[x, t'] \geq \nodeCost[x, \futt] = \nodeCost[u, \futt]
\]

By Lemma~\ref{lem:nodeCost is the optimal cost}, the optimal cost of travels to node $u$ arriving at time $\futt$ is $\nodeCost[u,\futt]$. Hence $\nodeCost[u, \futt]\leq \jcost(T') + \fcost(t_{x} - \futt)$. So we have, using the sub-additivity $\fcost$,
\begin{align*}
\nodeCost[u,\futt] + \fcost(\futt-t) 
&\leq \jcost(T') + \fcost(t_{x} - \futt) + \fcost(\futt-t) \\
&\leq \jcost(T') + \fcost(t_{x}-t) = \jcost(T)\leq \Const
\end{align*}
which contradicts Property~(\ref{eq:the contradiction with cost[i,t'] + fcost(t'-t)}).

\textbf{If $t' \geq t_{x}$, then,} when the pair $(x, t')$ was extracted from $\nodeCost$ (in Line~\ref{line:offline:extract min}), since an edge exists between $x$ and $u$ at time $t_{x}$, there is an iteration of the \emph{for} loop where $\pastt = t_{x}$ and we have
\[
\nodeCost[u,t_{x}] \leq \nodeCost[x,t'] + \fcost(t'-t_{x}) \leq \Const
\]
So $\nodeCost[u,t_{x}]$ is finite so, using Lemma~\ref{lem:nodeCost is the optimal cost}, $\nodeCost[u,t_{x}]=\jcost(T')$. We can use this to have:
\[
 \nodeCost[u,t_{x}] + \fcost(t_{x} - t) = \jcost(T') + \fcost(t_{x}-t) = \jcost(T)\leq \Const
\]
which contradicts Property~(\ref{eq:the contradiction with cost[i,t'] + fcost(t'-t)}).
\end{proofEnd}

\begin{textAtEnd}
Now we show that \texttt{ExtractTimeTravel} is correct and returns a solution to the problem, if it exists.
\end{textAtEnd}

\begin{theoremEnd}[confall,category=extract-prop]{lemma}
\label{lem:proof of extract}
Assuming $\nodeCost$ is constructed by Algorithm~\ref{alg:offline}, then \texttt{Extract\-Time\-Travel}$(u,t,\nodeCost, \pred)$ returns, if it exists, a travel with optimal cost to node $u$ arriving at time $t$.
\end{theoremEnd}

\begin{proofEnd}
By Lemma~\ref{lem:nodeCost is the optimal cost}, for any pair $(u,t)$, if $\nodeCost[u,t] < \infty$, then $\nodeCost[u,t]$ is the optimal cost of $\Const$-cost-constrained travels arriving at node $u$ and at time $t$.

We first prove the Lemma assuming $\nodeCost[u,t] < \infty$. By definition of $t'$ in Line~\ref{line to comute t' when nodecost in infinite} we know that recursive call to \texttt{ExtractTimeTravel} also satisfy this property.

We now show that the travel returned by \texttt{Extract\-Time\-Travel}$(u,t,$$\nodeCost)$ has cost $\nodeCost[u,t]$. We prove this result by induction on $(u,t)$  (in the order by which these pairs are extracted in Line~\ref{line:offline:extract min} of Algorithm~\ref{alg:offline}). If $u = \src$ the result is clear. Otherwise, we have by construction of \nodeCost:
\begin{equation}\label{eq:nodecost i-1 t' + fcost t'-t is nodecost i t}
\left\{
\begin{array}{rl}
     \nodeCost[v, t'] + \fcost(t'-t) &= \nodeCost[u, t] \\
     ((v,t'),(v,t),(u,t)) &= \pred[u, t]
\end{array}\right.
\end{equation}
for some $t'$ and some $v\in V$. Indeed, $\nodeCost[u, t]$ origins from $\nodeCost[v, t']$ either by extension on Line~\ref{algo1:line extract future 2} or on Line~\ref{algo1:line extract past 2} of Algorithm~\ref{alg:offline}. Observe that in the former case $t'<t$, but in this case $\fcost(t'-t)=0$, so the equation remains true.

Since $\nodeCost[v, t']$ is finite, and was extracted before $(u,t)$, then, by induction hypothesis, \\
\texttt{ExtractTimeTravel}$(v,t',\nodeCost)$ is a travel with optimal cost to $v$ arriving at time $t'$.
Hence we have that the cost of the returned travel is in fact
\begin{align*}
cost(\texttt{ExtractTimeTravel}(v,t',\nodeCost)\oplus(&(v,t'),(v,t),(u,t))) \\
&= \nodeCost[v,t']+\fcost(t'-t)
\end{align*}
This is equal to $\nodeCost[u,t]$, using Equation~\ref{eq:nodecost i-1 t' + fcost t'-t is nodecost i t}.

Now, assume $\nodeCost[u, t] = \infty$. If a cost-optimal $\Const$-cost-constrained travel $T$ exists to node $u$ arriving at time $t$, then $T$ must goes through a temporal edge $((x, u), t_1)$ for some $t_1\in \N$ and $x\in V$ \ie,
\[
T = T_1 \oplus ((u, t_1), (u, t))
\] 
and by construction $\nodeCost[u, t_1] = \jcost(T_1)$. So, when computing $t'$ in Line~\ref{line to comute t' when nodecost in infinite} we have 
\[
\nodeCost[u,t']+\fcost(t'-t) \leq \jcost(T)
\]
Then the recursive call of \texttt{ExtractTimeTravel} is made with $(u,t')$. Since we have $\nodeCost[u,t'] < \infty$, the previous property holds and the returned travel has the optimal cost towards $u$ arriving at $t'$, so its concatenation with $((u, t'), (u, t))$ is optimal to $u$ arriving at $t$.
\end{proofEnd}

\begin{theoremEnd}[conf]{theorem}\label{thm:algo offline is correct}
If the cost function $\fcost$ is in $\mathcal{UF}$, Algorithm~\ref{alg:offline} outputs a travel $T$ if and only if $T$ is a solution of the \cpbname problem.
\end{theoremEnd}

\begin{proofEnd}
By Lemma~\ref{lem:nodeCost can be used to find paths}, if there is a solution, $\tmin$ is the smallest time such that a travel exists to node $\dst$. Then, using $\nodeCost$, we can easily construct in a backward manner a solution using a \texttt{ExtractTimeTravel} procedure, as proved by Lemma~\ref{lem:proof of extract}
\end{proofEnd}

Let us now analyze the complexity of Algorithm~\ref{alg:offline}. We assume that retrieving the next or previous edge after or before a given time takes $O(1)$ time. For example, the graph can be stored as a dictionary that maps each node to an array that maps each time to the current, previous, and next temporal edges. This array can be made sparser with low complexity overhead to save space if few edges occur per time-instant.

Since each temporal edge is extracted from $\nodeCost$ at most once and the inner \emph{for} loop iterates over a subset of edges, the time complexity is polynomial in the number of temporal edges. We must also consider the time to extract the minimum from $\nodeCost$, which is also polynomial. If there are an infinite number of temporal edges\footnote{An evolving graph with an infinite number of edges can exist in practice even with bounded memory, e.g., when the graph is periodic.}, Lemma~\ref{lem:algo 1 terminates} shows that our algorithm always terminates, even if no solution exists. Therefore, its complexity is polynomial in the size of the finite subset of temporal edges extracted from $\nodeCost$.

\newcommand\EE{\mathcal{E}}
Let $\EE$ be the set of temporal edges $((u, v), t)$ such that $(u, t)$ or $(v,t)$ is extracted in Line~\ref{line:offline:extract min} of our algorithm during its execution. 

\begin{theorem}
If the cost function $\fcost$ is in $\mathcal{UF}$, then Algorithm~\ref{alg:offline} terminates in $O(|\EE|^2)$. 
\end{theorem}

\section{Offline $\Hmax$-history-constrained \pbname Algorithm}
\label{sec:offline_limited}

\SetKwFor{RepTimes}{repeat}{times}{end}
\begin{algorithm}[t]
\small
\caption{Offline $\Hmax$-history-constrained \pbname Algorithm}\label{alg:offline limited history}
\tcc{ \footnotesize $c[i, t-h, t]$ stores the cost of a cost optimal travel to node $x_i$, arriving before or at time $t-h$, that is $\Hmax$-history-constrained, and never reaches a time instant greater than $t$. $\qquad$ $\pred[u, t-h, t]$ stores the suffix of an optimal travel to $u$ arriving at $t-h$ that never reaches a time greater than $t$. }
$c[*] \leftarrow \infty;\qquad$
$c[\src, *] \leftarrow 0\qquad \pred[*] \leftarrow \bot$\label{line:algo H-constrained initialization}\;
$t_{\max} \leftarrow$ upper bound on the time reached by a cost-optimal travel to $\dst$\;
\For{$t = 0, 1, 2, \ldots, t_{\max}$}
{

\tcc{for simplicity, we assume $c[u, t-h, t] = \infty$ if $t-h < 0$}

\For{$u \in V$}
{
$c[u, t-h, t] \leftarrow \min\left(c[u, t-h, t-1], c[u, t-h-1, t-1]\right)$\;\label{line:the first min}
}

\RepTimes{$|V|$\label{line:repeat loop |V| times}}
{
\For{$u \in V$}
{
\For{$h = \Hmax, \Hmax - 1, \ldots, 0$}
{
$m \leftarrow \displaystyle\min_{\substack{t'\in[t-\Hmax, t]\\(u,v) \in E_{t'}}}\left(c[v, t', t] + \fcost(t' - (t-h))\right)$\;
\If{$c[u, t-h, t] < m$}
{
    $c[u, t-h, t] \leftarrow m$\;
     $\pred[u, t-h, t] \leftarrow (v, t')$ (with the corresponding min arguments)\;
}
}
}
}
\If{\textnormal{the minimum time instant $\tmin$  such that $c[\dst,\tmin,\tmin+\Hmax] < \infty$ exists}}{
\Return \texttt{ExtractHistoryConstrainedTravel}$(\dst, \tmin,\tmin+\Hmax, c)$\;
}
}
\Return $\bot$\;
\end{algorithm}

Section~\ref{sec:offline_unlimited} made the assumption that a given agent was able to go back to \emph{any} previous snapshot of the network. 
However, this hypothesis might not hold as the difficulty to go back in time may depend on how far in the future we already reach. Hence, we consider in this section that $\Hmax$ denotes the maximum number of time instants one agent can travel back to. In more detail, once an agent reaches time instant $t$, it cannot go back to $t' < t - \Hmax$, even after multiple jumps. 

\textbf{In this section, it is important to notice that the capability of BTT devices does not depend on the time when the agent uses it but rather on the largest time reached by the agent.} 

We present Algorithm~\ref{alg:offline limited history} that solve the $\Hmax$-history-constrained \pbname problem. The algorithm uses dynamic programming to store intermediary results. At each iteration, we update the optimal cost based on the best cost of previous nodes. For each node $x_i$ and time $t$ we need to store the best cost depending on the maximum time reached by the agent.

\begin{theoremEnd}[confall]{lemma}\label{lem:c function is the cost of an H-history-constrained optimal solution}
If the cost function $\fcost$ is in $\mathcal{UF}$, then, in Algorithm \ref{alg:offline limited history}, for any tuple $(u, t - h, t)$, if $c[u, t-h, t] < \infty$, then $c[u, t-h, t]$ is the cost of an $\Hmax$-history-constrained cost optimal travel towards node $u$, that arrives before or at time instant $t-h$, and never reaches a time instant greater than $t$.
\end{theoremEnd}
\begin{proofEnd}
We show by induction that, at the end of each iteration of the first \emph{for} loop, for each $t$, $c[u, t-h, t]$ is set to the cost of an $\Hmax$-history-constrained cost optimal travel towards node $x_i$, that arrives before or at time instant $t-h$, and never reaches a time instant greater than $t$.

It is clear that if $t=0$, the property is true because $c[\src, 0, 0]$ is initialized to 0, and cost 0 is propagated to all the nodes in the same connected component as $\src$.

Now assume that the property is true for any tuple smaller  $(u', t'-h', t')$ with $t' < t$.
Assume for the sake of contradiction that there exists a simple backward-cost optimal travel $T$ towards node $u$, arriving at time $t-h$ that is $\Hmax$-absolute-constrained, and never reaches a round greater than $t$, such that $cost(T) < c[i, t-h, t]$. 
We now consider three cases.

\textbf{If $T$ never reaches a time instant greater than $t-1$} (which implies in particular that $t\geq 1$), then, using the induction hypothesis on $T$, we have that $c[u, t-h, t-1] = \jcost(T)$. But by definition (Line~\ref{line:the first min}), we also have $c[u, t-h, t] \leq c[u, t-h, t-1]$, which is a contradiction (that would mean $c[u, t-h, t] \leq \jcost(T) < c[u, t-h, t]$).

\textbf{If $T$ arrives at $u$ before time $t$ \ie $T = T_0 \oplus ((u, t'),(u, t))$ with $t'< t$, and $T_0$ never reaches time instant greater than $t-1$}, then, using the induction hypothesis on $T_0$, we have that $c[u, t-h-1, t-1] = \jcost(T_0) = \jcost(T)$. But by definition (Line~\ref{line:the first min}), we also have $c[u, t-h, t] \leq c[u, t-h-1, t-1]$, which is a contradiction.

\textbf{Otherwise,} $T$ reaches time instant $t$ before arriving at $u$ \ie, $T = T' \oplus ((u_0, t'), (u_0, t)) \oplus T''$ where $t'<t$ and $T'$ is the longest prefix of $T$ that never reaches $t$. Thus, $T'$ arrives at $u_0$ before $t$ and never reaches $t$, so by induction hypothesis, 
\begin{equation}\label{eq:basecase until t_0}
    c[u_0, t, t] = c[u_0, t', t'] = \jcost(T').
\end{equation}

Moreover, 
\[
T'' = ((u_0,t = t_0),(u_1,t_0),(u_1,t_1),(u_2,t_1), \ldots, (u_k, t_{k-1}),(u_k = u, t_k = t))
\]
with $k<|V|$ because we assumed that $T$ is simple, $t_i \in [t-\Hmax, t]$ because the travel is $\Hmax$-history-constrained. Let $T_i$, $i\in[0,k]$, be the prefix of $T$ to $u_i$: 
\[
T_i = T'\oplus((u_0, t'), (u_0,t = t_0),(u_1,t_0),(u_1,t_1),(u_2,t_1), \ldots, (u_i, t_{i-1}),(u_i, t_i))
\]

Now, to reach a contradiction and prove the Lemma, we show by induction on $0\geq i<|V|$ that
\begin{equation}\label{eq:induction until t_i}
c[u_i, t_i, t] = \jcost\left(T_i\right)
\end{equation}
and that the value of $c[u_i, t_i, t]$ remains unchanged after the $i$-th iteration of the \emph{repeat} loop Line~\ref{line:repeat loop |V| times}.

The base case is given by Equation~(\ref{eq:basecase until t_0}).
Now we assume the property true for $i\in[0,k-1]$. At the $i+1$-th iteration of the \emph{repeat} loop we have 
\[
c[u_{i+1}, t_{i+1}, t] \leq c[u_i, t_i, t] + \fcost(t_{i} - t_{i+1}) = \jcost(T_{i+1})
\]
Moreover, $\jcost(T_{i+1}) \leq c[u_{i+1}, t_{i+1}, t]$, because $c[u_{i+1}, t_{i+1}, t]$ is the cost of a travel so it cannot be smaller than the cost of the cost-optimal travel $T_{i+1}$. So $c[u_{i+1}, t_{i+1}, t] = \jcost(T_{i+1})$ and its value never changes after the $i+1$-th iteration of the \emph{repeat} loop. This conclude the induction and the proof of the Lemma.
\end{proofEnd}

\begin{theoremEnd}[confall,category=extract-prop-2]{lemma}\label{lem:proof of extract history}
Assuming $c$ is constructed by Algorithm~\ref{alg:offline limited history} and $c[u, t, t+\Hmax] < \infty$, then \texttt{ExtractHistoryConstrainedTravel}$(u,t, t+\Hmax,c)$ returns an $\Hmax$-history-constrained travel with optimal cost to node $u$ arriving at time $t$.
\end{theoremEnd}
\begin{proofEnd}
First, we show that the number of recursive call to \texttt{Extract\-History\-Constrained\-Travel} is finite. 
Either the function is called with a value $t_M' < t_M$ or with the same value $t_M$ but a different pair $(v, t') \neq (u, t)$. So we have to show that the number of recursive calls with the same value $t_M$ is finite. 
Every time \texttt{ExtractHistoryConstrainedTravel} is called with the same value $t_M$, we have that $(v, t') = \pred[u, t, t_M]$ such that $c[v,t', t_M]\leq c[u,t, t_M]$.
Assume we have $k+1$ node $u_0, u_1,\ldots, u_k$ such that $\pred[u_i, t_i, t_M] = (u_{i-1}, t_{i-1})$, for all $i\in[1, k]$. Then we have $c[u_0,t_0, t_M]\leq c[u_i,t_i, t_M]$. In order to create a loop, $\pred[u_0, t_0, t_M]$ must be updated to be $(u_i, t_i)$ for some $i$. For this to happen, $c[u_0, t_0, t_M]$ must be strictly smaller than $c[u_i, t_i, t_M]$, which is a contradiction. Since no loop can occur, \texttt{ExtractHistoryConstrainedTravel} is called only a finite number of times.

Now we prove the Lemma. Recall that, for any tuple $(u, t, t_M)$, if $c[u, t, t_M] < \infty$, then $c[u, t, t_M]$ is the cost of an $\Hmax$-history-constrained cost optimal travel towards node $u$, that arrives before or at time instant $t$, and never reaches a time instant greater than $t_M$. So it is enough to prove that the travel returned by \texttt{ExtractHistoryConstrainedTravel}$(u,t,t_M,c)$ has cost $c[u, t, t_M]$, never reaches a time greater than $t_M$ and is $\Hmax$-history-constrained. We prove this result by induction on the number $r$ of recursive calls to \texttt{ExtractHistory\-Constrained\-Travel}. If $r = 0$ the result is clear because $u=\src$.
Otherwise, we have by construction of $c$, two cases.

\textbf{If $\pred[u, t, t_M] = \bot$}, then the value is never changed after Line~\ref{line:the first min}. So we know there exists $t'$ and $t_M'<t_M$ such that $c[u, t, t_M] = c[u, t', t_M']$. By inductive hypothesis, \texttt{ExtractHistoryConstrainedTravel}$(u, t', t_M')$ is a cost-optimal travel to $u$ arriving at time $t'$ that never reaches $t_M'$. If $t' < t$, we obtain a cost-optimal travel arriving at time $t$ by concatenating $((u,t'),(u, t))$.

\textbf{If $\pred[u, t, t_M] = (v, t')$}, then the travel return by the call to \texttt{Extract\-History\-Constrained\-Travel}$(v, t', t_M)$ is cost-optimal towards $v$ arriving at time $t'$ and its concatenation with $((v,t'), (u,t'), (u,t))$ is cost-optimal towards $u$ arriving at time $t$ because by construction of $c$, $c[v,t',t_M] + \fcost(t'-t)$, which is the cost of the concatenation, is exactly $c[v,t,t_M]$, which the optimal cost.

\end{proofEnd}

\begin{theoremEnd}[conf]{theorem}
If the cost function $\fcost$ is in $\mathcal{UF}$, then Algorithm \ref{alg:offline limited history} solves the  $\Hmax$-history-constrained \pbname problem and has $O(n^2\Hmax(\tmin+\Hmax))$ complexity, with $\tmin$ the delay of a solution.
\end{theoremEnd}

\begin{proofEnd}
If no solution exists, since $t_{\max}$ is finite, our algorithm returns $\bot$. Otherwise, using Lemma~\ref{lem:c function is the cost of an H-history-constrained optimal solution} and Lemma~\ref{lem:proof of extract history} we know that Algorithm~\ref{alg:offline limited history} returns a solution to the $\Hmax$-history-constrained \pbname problem.

Regarding the complexity, computing $t_{\max}$ can be done in $O(n^2)$ using a modified version of Algorithm~\ref{alg:offline}, were at most on temporal edge is extend per neighbor (the one with the greatest time).

Then, we exit the main loop after reaching $t = \tmin + \Hmax$, and the three inner loops have complexity $O(n^2\Hmax)$, so the complexity of the algorithm is in $O(n^2\Hmax(\tmin + \Hmax))$.
\end{proofEnd}

\begin{textAtEnd}[category=extract-prop-algo]
\begin{algorithm}[h]
\small
\SetAlgoRefName{{ExtractHistoryConstrainedTravel}}
\caption{Extract an $\Hmax$-history-constrained \pbname travel to the given destination}
\SetKwInOut{Input}{input}
\Input{$u\in V,\quad t\in\N,\quad t_M\in\N,\quad c$}
\If{$u=\src$}{
    \Return $((\src,0), (\src, t))$\\
}
\If{$c[u,t,t_M]=\infty$}{
    \Return $\bot$\\
}
\If{$\pred[u, t, t_M] \neq \bot$}{
    $(v, t') = \pred[u, t, t_M]$\;
    \Return \texttt{ExtractHistoryConstrainedTravel}($v, t', t_M,c$)$\oplus((v,t'), (u, t'), (u,t))$\\
}
Let $t$ and $t_M' < t_M$ such that $c[u,t,t_M] = c[u, t', t_M']$
\If{$t = t'$}
{
\Return \texttt{ExtractHistoryConstrainedTravel}($u,t', t_M',c$)\;
}
\Return \texttt{ExtractHistoryConstrainedTravel}($u,t', t_M',c$)$ \oplus((u,t'),(u,t))$\;

\end{algorithm}
\end{textAtEnd}

\section{Conclusion}

We presented the first solutions to the optimal delay optimal cost space-time constrained travel planning problem in dynamic networks, and demonstrated that the problem can be solved in polynomial time, even in the case when backward time jumps can be made up to a constant, for any sensible pricing policy. It would be interesting to investigate the online version of the problem, when the future of the evolving graph is unknown to the algorithm.

\bibliographystyle{splncs04}
\bibliography{biblio}

\ifreport
\else

\fi
\end{document}